\documentclass[journal=jctcce,manuscript=article,layout=twocolumn]{achemso}

\usepackage{amssymb}
\usepackage[utf8]{inputenc}
\usepackage[version=3]{mhchem} 
\usepackage{hyperref}
\usepackage{braket}
\usepackage{bm}
\usepackage[svgnames]{xcolor}
\usepackage{natbib}
\usepackage{graphicx, nicefrac, textcomp}
\usepackage{soul}
\usepackage{amsmath}
\usepackage{booktabs}

\usepackage{algorithm}
\usepackage{algpseudocode}
\usepackage{float}
\usepackage{enumitem}

\usepackage[authormarkuptext=name,authormarkup=none]{changes}
\definecolor{red}{RGB}{255, 0, 0}
\definecolor{blue}{RGB}{0, 0, 255}
\definecolor{green}{RGB}{0, 192, 0}
\definechangesauthor[name={Yorick}, color={blue}]{YLAS}
\definechangesauthor[name={Gianluca}, color={cyan}]{GL}

\title{A Neural-Network-Based Selective Configuration Interaction Approach to Molecular Electronic Structure}

\author{Yorick L.\ A.\ Schmerwitz}
\email{schmerwitz@kofo.mpg.de}
\affiliation[1]{Science Institute and Faculty of Physical Sciences, University of Iceland, 107 Reykjav\'ik, Iceland}
\alsoaffiliation[4]{Max-Planck-Institut f\"ur Kohlenforschung, 45470 M\"ulheim an der Ruhr, Germany}
\author{Louis Thirion}
\affiliation[2]{Department of Physics, Friedrich-Alexander-Universit\"at Erlangen/N\"urnberg, 91058 Erlangen, Germany}
\alsoaffiliation[1]{Science Institute and Faculty of Physical Sciences, University of Iceland, 107 Reykjav\'ik, Iceland}
\author{Gianluca Levi}
\affiliation[1]{Science Institute and Faculty of Physical Sciences, University of Iceland, 107 Reykjav\'ik, Iceland}
\author{Elvar~\"O.~J\'onsson}
\affiliation[1]{Science Institute and Faculty of Physical Sciences, University of Iceland, 107 Reykjav\'ik, Iceland}
\author{Pavlo Bilous}
\affiliation[3]{Max Planck Institute for the Science of Light, Staudtstraße 2, 91058 Erlangen, Germany}
\author{Hannes J\'onsson}
\email{hannes_jonsson@brown.edu}
\affiliation[1]{Science Institute and Faculty of Physical Sciences, University of Iceland, 107 Reykjav\'ik, Iceland}
\author{Philipp Hansmann}
\email{philipp.hansmann@fau.de}
\affiliation[2]{Department of Physics, Friedrich-Alexander-Universit\"at Erlangen/N\"urnberg, 91058 Erlangen, Germany}
\alsoaffiliation[1]{Science Institute and Faculty of Physical Sciences, University of Iceland, 107 Reykjav\'ik, Iceland}
\begin{document}


\begin{abstract}
By combining Hartree-Fock with a neural-network-supported quantum-cluster solver proposed recently in the context of solid-state lattice models, we formulate a scheme for selective neural-network configuration interaction (NNCI) calculations and implement it with various options for the type of basis set and boundary conditions. The method's performance is evaluated in studies of several small molecules as a step toward calculations of larger systems. In particular, the correlation energy in the N$_2$ molecule is compared with published full CI calculations that included nearly $10^{10}$ Slater determinants, and the results are reproduced with only $4\cdot10^{5}$ determinants using NNCI. A clear advantage is seen from increasing the set of orbitals included rather than approaching full CI for a smaller set. The method's high efficiency and implementation in a condensed matter simulation software expands the applicability of CI calculations to a wider range of problems, even extended systems through an embedding approach.
 
\end{abstract}

\maketitle

\section{Introduction}
\label{section:intro}

The accurate determination of the electronic ground state of atomic scale systems, such as molecules and condensed matter, presents a critical challenge primarily due to the exponential growth of the Hilbert space with the number of electrons and orbitals involved. For molecules, obtaining a precise estimate of the energy of the ground electronic state is crucial for understanding bonding characteristics, structure, and reaction pathways. Full configuration interaction (FCI) methods are among the most direct approaches to solving the electronic Schrödinger equation. 
There, the electronic wave function is expressed as a linear combination of basis vectors, namely Slater determinants (SDets). Determining the eigenvectors of the many-body Hamiltonian in such a basis allows, in principle, for the exact treatment of electron correlations \cite{Szabo_ModnQChem_2012}.\\

However, the practical application of FCI is limited by the ``exponential wall", i.e.\ the combinatorial growth of the number of basis SDets with increasing system size. Even for a small diatomic molecule, such as N$_2$, a remarkably large number of SDets is needed, pushing the traditional computational approaches to their limits \cite{ROSSI1999}. One strategy for circumventing the explosive dimensional increase is the so-called \emph{embedding technique}. Here, a system that is too large for full computation is subdivided into a smaller strongly correlated part, the embedded cluster, which is treated with a high-level, quantum many-body technique (as, e.g., CI) while the rest is treated at a more approximate, mean-field level using Hartree–Fock \cite{Szabo_ModnQChem_2012} or Kohn–Sham density functional theory \cite{Kohn1965, Hohenberg1964}. Examples in quantum chemistry are embedded correlated wave function schemes and embedding potential schemes \cite{Libisch2014,Jones2020}. While traditional and newly developed embedding techniques show great potential, the size of the embedded cluster can still be the prohibiting factor for capturing essential quantum mechanical correlations. To this end, based on the observation that usually only a small subset of the SDets contributes significantly to the description of the eigenfunctions \cite{Ivanic2001}, a large variety of selected CI methods has been developed, see for example Refs.~\citenum{Huron1973,Greer1998, Garniron2018,Tubman2020} and references therein.
Alternatively, a Monte Carlo method ansatz has been proposed where important configurations are stochastically sampled, as has been implemented in, e.g., FCIQMC\cite{FCIQMC}.
\\

The advent of machine learning (ML) algorithms has given a new twist to these efforts. ML techniques have been applied in the context of selective CI computations and have already shown great potential \cite{Coe_MLCI_JChemTC_2018,Coe_JChemTC_2019,Jeong_ALCI_JChemTC_2021, Chembot, RLCI, Herzog2023,Molchanov2022,MLGRASP, Bilous_PhysRevA.110.042818,Bilous2024}. This includes selection of relevant SDets based on a regression neural network (NN) applied to the expansion coefficients \cite{Coe_MLCI_JChemTC_2018,Coe_JChemTC_2019}, and schemes which leverage NN-based classification to distinguish ``important" from ``unimportant" SDets directly \cite{Jeong_ALCI_JChemTC_2021}. Related approaches have recently been developed also for the structure and dynamics of light nuclei \cite{Molchanov2022}, for accurate calculations of energy levels of isolated atoms and ions\cite{MLGRASP, Bilous_PhysRevA.110.042818}, and most recently for strongly correlated solids in the context of effective models \cite{Bilous2024}.\\

In this work, we apply the NN-supported iterative approach from Ref.~\citenum{Bilous2024}, which employs a convolutional NN classifier in an \emph{active learning} scheme. In contrast to supervised learning, the data used in active learning for the NN training are generated in each iteration on the fly. In turn, the trained NN controls the basis of SDets used in each iteration within a parameter-free CI calculation. In this way, we demonstrate a post-Hartree-Fock, selective CI framework, which represents a stepping stone toward the implementation of more advanced embedding schemes as needed, e.g., for the simulation of heterogeneous catalysis and optical excitation of defects in solids, where strongly correlated active sites interact with an extended environment that can be treated on a mean-field level.

In the following, we present the results of benchmark calculations of molecules and report the correlation energy, binding energy, and dissociation curve of the paradigmatic N$_2$ molecule, as well as the correlation energy of the
H$_2$O, NH$_3$, and CO molecules. Our findings show that our NN-based approach works well in boosting the CI efficiency for material-realistic calculations involving complex multi-orbital two-particle interactions. Indeed, we find the NNCI method to leverage the increased dimensions of the configuration space rather than being hindered by its combinatorial growth.

\section{Methods}
For the computation of the many-body wave function of a molecule, we can cast the problem into a Hamiltonian form and write
\begin{equation}
\label{eq:Hfull}
H=H^0 +H^\mathrm{int} - \mathrm{MF}\left[H^\mathrm{int}\right]
\end{equation}
with 
\begin{align}
H^{0} &= \sum_{i,j,\sigma} \; t_{ij}  \;c^\dagger_{i,\sigma} c_{j,\sigma}^{\phantom{\dagger}}\\
H^\mathrm{int} &= \sum_{\substack{i,j,k,l \\ \sigma,\sigma'}} U_{ijkl} \;c^\dagger_{i,\sigma} c_{j,\sigma}^{\phantom{\dagger}} c^\dagger_{k,\sigma'} c_{l,\sigma'}^{\phantom{\dagger}}
\end{align}
where $c^\dagger_{i,\sigma}$ and $c_{i,\sigma}$ are fermionic creation and annihilation field operators, respectively, with orbital indices $i$ and spin indices $\sigma$. $t_{ij}$ and $U_{ijkl}$ are the single- and two-particle integrals
\begin{align}
t_{ij} &\equiv \int \text{d}\mathbf{r} \, \psi^*_i(\mathbf{r}) \left(-\frac{1}{2}\nabla^2 + V^\text{eff}(\mathbf{r}) \right) \psi_j(\mathbf{r})\\
U_{ijkl} &\equiv \int \text{d}\mathbf{r}\,\text{d}\mathbf{r'} \,\psi^*_i(\mathbf{r}) \psi^*_j(\mathbf{r'}) \frac{1}{|\mathbf{r}-\mathbf{r'}|}  \psi_k(\mathbf{r}) \psi_l(\mathbf{r'})
\end{align}
where $V_\text{eff}(\mathbf{r})$ is the self-consistent mean-field potential, and the integrals are evaluated on the Hartree-Fock eigenbasis of molecular orbitals (MOs) $\psi_i(\mathbf{r})$. The term $\mathrm{MF}\left[H^\mathrm{int}\right]$ in Eq.~\eqref{eq:Hfull} is the mean-field decoupled interaction operator. Since its contributions are implicitly included in the Hartree-Fock single particle integrals $t_{ij}$, we need to subtract it to avoid double counting (see Appendix~\ref{app:DC}). 

We calculate the ground state of $H$ in two steps. In the first step, the integrals $t_{ij}$ and $U_{ijkl}$ are calculated. In the second step, the ground state of $H$ 
is evaluated in a selected many-body basis using the NN-based algorithm. All calculations in this work are performed retaining spin symmetry.

\subsection{Hartree-Fock}
\label{sec:hf}
The occupied MOs used to compute the single- and two-particle integrals, $t_{ij}$ and $U_{ijkl}$\,, are the canonical orbitals obtained in Hartree-Fock calculations using a development version of the GPAW software \cite{GPAW2024}. 
The GPAW software provides real space grid, plane wave, and localized atomic basis set compatibility. Calculations can be performed in either of these frameworks and 
in combination, e.g., when a plane wave computation makes use of the real space grid when beneficial.
The projector augmented wave (PAW) method \cite{paw1,paw2} is used to treat the electrons near the nuclei, and the core electrons for each atom are frozen to the result of a reference scalar relativistic calculation of the isolated atom. 
The smooth pseudo wave functions for the occupied MOs containing valence electrons are described here using a plane wave basis set. 
The Hartree-Fock calculations are carried out using a direct minimization approach \cite{dm1,dm2} employing the L-BFGS algorithm.
The initial guess orbitals for the energy minimization are constructed using a set of numerical atomic orbitals centered at the positions of the nuclei\cite{Larsen2009}. This set contains two sets of numerical $s$, and one set of numerical $p$ orbitals for the hydrogen atom, and four sets of numerical $s$, four sets of numerical $p$, and two sets of numerical $d$ orbitals for the nitrogen and oxygen atoms, with the set for the carbon atom having one less set of numerical $p$ orbitals. 
The atomic orbital sets are first evaluated on a real space grid and subsequently, Fourier transformed to the plane wave representation.
The calculations are considered converged when the squared residual of the Hartree-Fock equations is below 10$^{-11}\,\text{eV}^{2}$ per valence electron for the optimal orbitals that minimize the Hartree-Fock energy. Since the virtual orbitals do not impact the energy, they are not optimized and correspond to the initial guess orbitals, adjusted only to ensure orthonormality. In the frozen core approximation used here, the 1$s$ orbitals of second-row atoms are not optimized either, and these core orbitals are also not considered in the subsequent CI calculations. The calculation is performed for the maximal number of MOs. Smaller numbers of orbitals for the CI computations are generated by systematically omitting the highest-energy virtual orbital sets. All Hartree-Fock calculations are performed with a grid spacing of $0.18\,\text{\AA}$ and a plane wave energy cutoff of $1000$\,eV. The size of the cubic simulation cell is chosen such that at least 12.5\,\AA\ of vacuum are included between each atom and the closest cell boundary.
Further details on the Hartree-Fock calculations are provided in Appendices \ref{app:HF_single} and \ref{app:HF_pair}.

\subsection{NN-supported CI}

\paragraph{Selective CI -}
In the CI framework, the exact many-body ground state which fulfills 
\begin{equation}
H  \ket{\Psi^\text{ex}_\text{gs}} = E^\text{ex}_\text{gs}  \ket{\Psi^\text{ex}_\text{gs}}
\end{equation}
for a given Hamiltonian (Eq.~\ref{eq:Hfull}) is typically expanded in SDets $\ket{\phi}$ forming an orthonormal basis $\{\phi\}$ of the full Hilbert space ${\cal H}^\text{full}=\text{span}\left(\{ \phi \}\right)$:
\begin{equation}
\ket{\Psi^\text{ex}_\text{gs}} = \sum_{{\cal H}^\text{full}} c_\phi \ket{\phi}\,.
\end{equation}
In selective CI, we aim to approximate the exact ground state by searching for the most relevant subspace of ${\cal H}^s \subset {\cal H}^\text{full}$ with $\text{dim}\left({\cal H}^s\right) \ll \text{dim}\left({\cal H}^\text{full}\right)$, so that

\begin{equation}
\label{eq:psi_approx}
\ket{\Psi^\text{ex}_\text{gs}} \approx \ket{\Psi_\text{gs}} = \sum_{{\cal H}^\text{s}} c_\phi \ket{\phi}\,.
\end{equation}

An iterative protocol for a selective CI procedure typically starts from a small initial basis $\{\phi^\text{init}\}$. The basis can then be iteratively extended by an extension operator $\hat{\cal O}$ which, by acting on determinants of a given set, generates new configurations (given that the set does not span an eigenspace of the operator). In each iteration, the Hamiltonian eigenproblem is solved on the corresponding subspace yielding the approximate ground state wave function and energy. Due to the combinatorial growth in the number of generated configurations, this procedure quickly becomes computationally infeasible. Consequently, a selection protocol must be introduced at an appropriate stage. In this work, we use the NN-based selection from Ref.~\citenum{Bilous2024} detailed below. Benchmarks from this prior study demonstrate that NN-supported selection for the many-body basis significantly outperforms simpler truncation schemes, such as a simple energy cutoff.

\paragraph{NNCI iterations -}
We summarize an iteration of our NN-supported selection process in the chart Algorithm~\ref{alg:NN-CI} which is applied once the extension procedure leads to an intractably large basis set $\{\phi\}$. The procedure selects determinants based on their importance, defined by whether their CI expansion coefficients exceed a given cutoff. However, it is not the cutoff itself but rather the target number of determinants to be appended at each step that serves as the primary control parameter. The cutoff is adjusted accordingly in each iteration\cite{Bilous2024}. The size of the random selection drawn at the beginning of Algorithm~\ref{alg:NN-CI} is the second control parameter.
\begin{algorithm}[H]
\caption{NNCI iteration}\label{alg:NN-CI}
\begin{algorithmic}[1]
\State\textbf{Input:}
\begin{itemize}
    \item SDets $\{\phi\}$ currently in CI expansion 
    \item Pool of candidate SDets $\{\phi^{\text{pool}}\} = \hat{\cal O}\{\phi\} \setminus \{\phi\}$
    \item Number of most important SDets to include from the pool
    \item Size of the random selection
\end{itemize}
\State Take random sample $\{\phi^{\text{rand}}\} \subset \{\phi^{\text{pool}}\}$
\State Find ground state on $\{\phi\} \cup \{\phi^{\text{rand}}\}$
\State Adjust the cutoff parameter to split $\{\phi^{\text{rand}}\}$ in important $\{\phi^{\text{rand}}_{\text{impt}}\}$ and unimportant $\{\phi^{\text{rand}}_{\text{unimpt}}\}$ classes 
\State Train NN classifier on $\{\phi\} \cup\{\phi^{\text{rand}}\}$
\State Predict important SDets $\{\phi^{\text{NN}}_{\text{impt}}\} \subset \{\phi^{\text{pool}}\}$ using trained NN
\State Find ground state on
    $\{\phi\} \cup \{\phi^{\text{NN}}_{\text{impt}}\} \cup \{\phi^{\text{rand}}_{\text{impt}}\}$
\State Eliminate false positives from $\{\phi^{\text{NN}}_{\text{impt}}\}$
\State \textbf{Output:} $\{\phi\} \leftarrow \{\phi\} \cup \{\phi^{\text{NN}}_{\text{impt}}\} \cup \{\phi^{\text{rand}}_{\text{impt}}\}$
\end{algorithmic}
\end{algorithm}
In this way, the CI basis is iteratively appended with SDets which were classified to be important by the NN. It is noteworthy that after the selection described in Algorithm~\ref{alg:NN-CI} is performed in a given iteration, observables can be computed on the obtained basis $\{\phi\}$ for monitoring the convergence. While in the present case, we have focused on the ground state energy, the NNCI scheme can be generalized straightforwardly to converge the energy of excited states or other quantities, such as magnetic moments or orbital occupations simultaneously.

From the ML perspective, this approach belongs to the \emph{active learning} paradigm, which involves decision-making based on repetitive interaction with the ``environment" and learning optimal actions from the obtained data \emph{on-the-fly}, a strategy shared with reinforcement learning. However, the learning is not driven here by obtained rewards but is performed directly on the generated data in a supervised manner.

\paragraph{Implementation in the present work -}

For the calculations presented in this work, we use the Hartree-Fock single-SDet ground state as an initial basis ($\{\phi^\text{init}\}=\left\{\phi^\text{HF}_\text{gs}\right\}$) and the dominating excitations from the full Hamiltonian \eqref{eq:Hfull} (with amplitudes larger than $0.02$) as the extension operator $\hat{\cal O}$. Before the NN selection procedure is started, we extend twice without dropping any determinants. In this way, we always include the most important configurations up to quadruple excitations (concerning the Hartree-Fock ground state).
Higher order excitations are then reached in the subsequent ML-iterations and the corresponding SDets are presented to our NN classifier to be either selected or discarded as we describe above.

\begin{figure}[t]
     \begin{center}
     \includegraphics[width=\columnwidth]{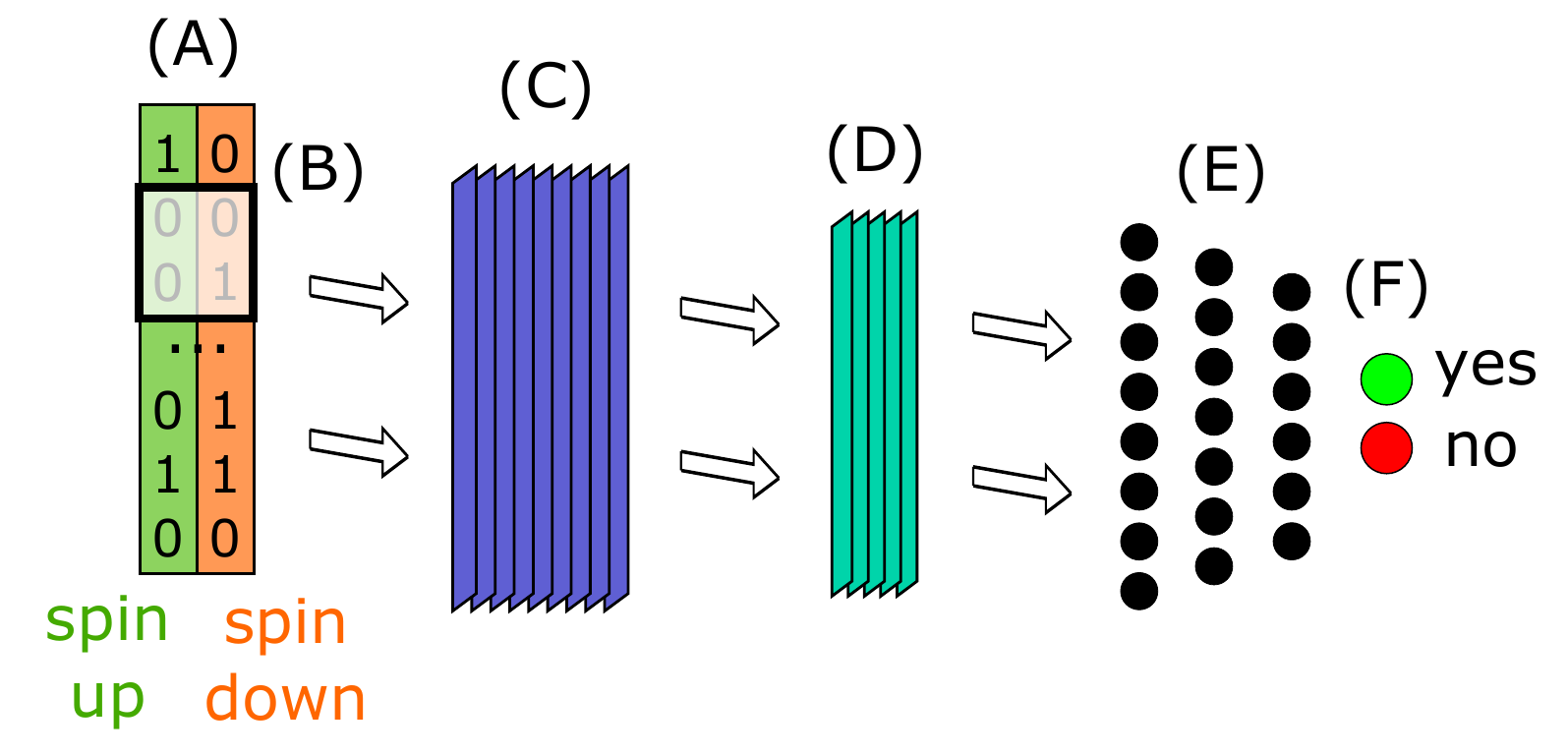}
     \end{center}
     \caption{
     The architecture of the convolutional NN used in this work. A candidate SDet $\ket{\phi}$ is given as input (A) and eventually classified as important or unimportant (F). The data propagate through the convolutional block (A)---(D) and the dense block (E)---(F).}
     \label{fig:nn_architecture}
\end{figure}

The architecture of our convolutional NN classifier \cite{Goodfellow2016} is shown schematically in Fig.~\ref{fig:nn_architecture}. The parts of the convolutional block (A)---(D) and the dense block (E)---(F), as well as other NN-related and implementation details are presented in Appendix~\ref{app:NN}. It has been demonstrated in similar computations for energy levels of isolated atoms\cite{MLGRASP} that such an NN of the convolutional type yields more stable results than usual dense NNs.

The NNCI computations described here are performed using SOLAX \cite{SOLAX}, a Python framework for solving the CI problem for fermionic quantum systems with neural network support.

%

\paragraph{Relation to other ML algorithms -} 
An iterative active-learning approach as the one described above was first proposed in Refs.~\citenum{Coe_MLCI_JChemTC_2018, Coe_JChemTC_2019} in which the NN, however, was employed as a regressor for the prediction of transformed CI coefficients. Later, direct classification approaches were introduced\cite{Jeong_ALCI_JChemTC_2021, Chembot}, sorting determinants into ``important" and ``unimportant" classes based on a cutoff parameter.
In the present work, a softmax NN classifier is used, as in the recent related works~\citenum{MLGRASP, Bilous_PhysRevA.110.042818, Bilous2024}. 
This is today a standard approach for solving general classification tasks such as in image recognition. 

Improvements to this scheme were first introduced in Ref.~\citenum{Jeong_ALCI_JChemTC_2021} where an iteration-dependent, running cutoff was used to mitigate false convergence to cutoff-dependent energy. Subsequently, Ref.~\citenum{Bilous2024} further refined this approach by automatically computing the running cutoff based on a desired fraction of important determinants to be added in each iteration. Moreover, the work of Ref.~\citenum{MLGRASP} demonstrated that an NN of the convolutional architecture yields more stable results than usual dense NNs when applied to the computation of energy levels of isolated atoms. In the present work, we also employ a convolutional NN classifier.
As shown below, our approach proves particularly effective in handling  high-dimensional configuration spaces. This allows us to push the boundaries of computational feasibility. In the case of N$_2$, e.g., we utilize up to 52 MOs, showcasing the method’s ability to manage very large configuration spaces efficiently while maintaining high accuracy.

The ML approach used here belongs to the branch of \emph{discriminative} ML algorithms, which aim at unveiling the dependence between an input $X$ and the ``correct'' output $y$ based on the training set, to predict $y$ for yet unseen $X$. A complementary paradigm is the \emph{generative} approach, which has the purpose of learning the probability distribution of $p(X|y)$ for each $y$, which allows to sample new data $X$ with the proper label $y$. In the domain of NN-supported selective CI, this alternative route has been taken by Herzog {\it et al.} \cite{Herzog2023}. In their \emph{CIgen} method, a restricted Boltzmann machine (RBM) is employed as a generative NN setting, to expand the configuration space iteratively by proposing new determinants that contribute significantly to the wave function.

It should be noted that there are also strategies to employ deep-learning techniques by representing the wave function in continuous space directly through an NN as implemented in PauliNet\cite{PauliNet1, PauliNet2} (instead of expanding the many-body wave function in a discrete configuration space). The fundamental differences in the formalism complicate direct benchmarking of the numerical performance/efficiency. Instead, comparisons should be made in terms of the accuracy in reproducing experimentally measurable observables, such as molecular and solid-state structures, energy and intensity of spectroscopic transitions, magnetic properties, and others.

\paragraph{Computation of the energy -} 
For the application of the NNCI scheme to the ground state, the Hartree-Fock, total, and correlation energy are evaluated based on Hamiltonian \eqref{eq:Hfull} as
\begin{align}
\label{eq:energies}
E^\text{HF}_\text{gs}&\equiv\braket{\phi^\text{HF}_\text{gs}| H | \phi^\text{HF}_\text{gs}}\,,\\
E_\text{gs}&\equiv\braket{\Psi_\text{gs}| H | \Psi_\text{gs}}\,,\\
E_\text{corr}&\equiv  E_\text{gs} - E^\text{HF}_\text{gs}\,,
\end{align}
respectively. Here, $\ket{\phi^\text{HF}_\text{gs}}$ is the spin-restricted Hartree-Fock (RHF) single-SDet ground state and $\ket{\Psi_\text{gs}}$ the approximate many-body ground state \ref{eq:psi_approx}. Moreover, for the N$_2$ molecule, we perform computations for various values of the fixed distance $a$ between the two nitrogen atoms up to $a=3.0$\,\AA. The resulting dissociation curve $E_\text{gs}(a)$ allows for the calculation of the binding energy
\begin{equation}
\label{eq:Ebind}
    E_\text{bind}\equiv E_\text{gs}(a\rightarrow\infty)-E_\text{gs}(a_\mathrm{min})\,,
\end{equation}
where $a_\mathrm{min}$ is the location of the minimum and the assumption is that $E_\text{gs}(\infty)\approx E_\text{gs}(a=3.0\,\text{\AA})$.

\section{Results}

\begin{figure}[t]
    \centering
    \includegraphics[width=0.48\textwidth]{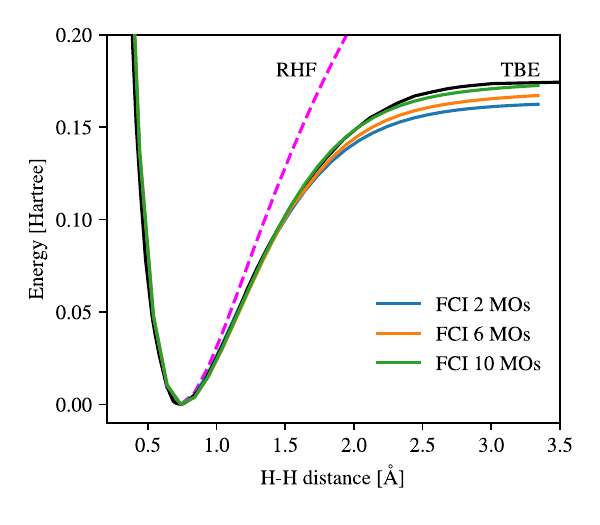} 
    \caption{H$_2$ dissociation curve obtained using spin-restricted Hartree-Fock (RHF) and FCI   
     with different number of molecular orbitals (MOs).
    In each case, the minimum energy is set to E=0. 
    The theoretical best estimate (TBE)\cite{Kolos1965} is shown for comparison.
    \label{fig:CurveH2}}
\end{figure}

\paragraph{H$_2$ benchmark -} 
Before applying the NNCI method to larger molecules, we first validate our computation protocol using the dissociation curve of the hydrogen molecule as a benchmark since the smaller Hilbert space allows for FCI calculations without NN assistance. In Fig.~\ref{fig:CurveH2}, the computed dissociation curve of H$_2$ is shown along with a comparison to the theoretical best estimate \cite{Kolos1965}. 
 
Specifically, we show the dependency of the FCI computation concerning the size of the underlying Hartree-Fock (single-particle, MO) basis. A basis of ten MOs (corresponding to a many-body Hilbert space spanned by $\binom{2\cdot10}{2} =$ 190 SDets) is sufficient to converge reasonably close to the exact curve and proves the feasibility of the approach presented here.

\begin{figure}[t]
    \centering
    \includegraphics[width=0.48\textwidth]{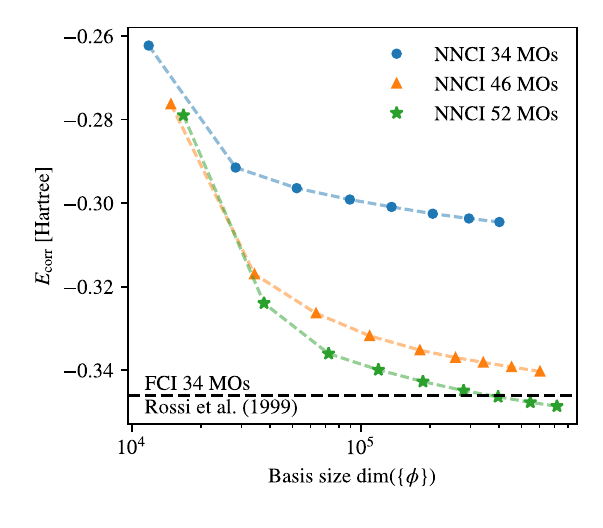} 
    \caption{N$_2$ correlation energy computed with NNCI as a function of the number of SDets spanning the many-body Hilbert space. Data for three different sets of MOs are shown. The benchmark (black dashed line) shows results from the FCI calculation in Ref. \citenum{ROSSI1999} using 34 MOs, which corresponds to approximately $10^{10}$ (symmetry adapted) basis SDets. The calculations were done for $a=2.1\,a_0\approx1.111\,\mathrm{\AA}$.}
    \label{fig:Ecorr}
\end{figure}

\begin{figure*}[t]
    \centering
    \includegraphics[width=0.98\textwidth]{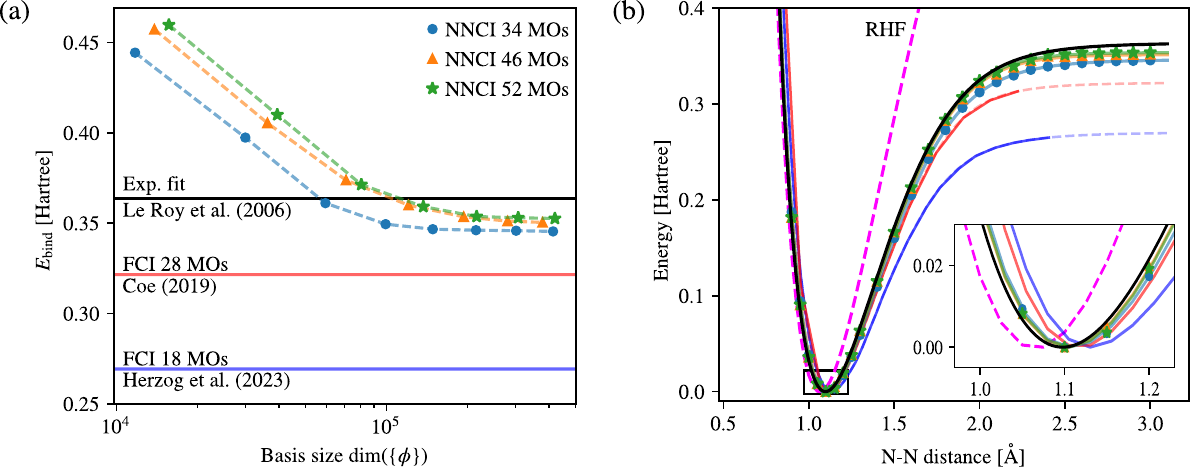} 
    \caption{
Results of NNCI calculations of an N$_2$ molecule using three different MO basis sets, and comparison with fitted energy curved obtained by fitting experimental data\cite{N2exp}, as well as FCI calculations using 18 (Ref. \citenum{Herzog2023}), and 28 (Ref. \citenum{Coe_JChemTC_2019}) MOs. (a) Binding energy of the N$_2$ molecule as a function of the number of SDets spanning the many-body Hilbert space. (b) Dissociation curves with an additional reference to spin-restricted Hartree-Fock (RHF). The curves have been aligned at the minimum energy $E_\text{gs}(a_\text{min})\equiv 0$. The inset shows 
the region
around the energy minima.}
    \label{fig:Ebind_diss}
\end{figure*}

\paragraph{N$_2$ Correlation energy.}
To demonstrate the accuracy of the methodology by comparing to an FCI benchmark\cite{ROSSI1999}, we first compute the correlation energy $E_\text{corr}$ (see eq.~\ref{eq:energies}) at the same distance as Ref.~\citenum{ROSSI1999} of $a=2.1\,a_0\approx1.111\,\mathrm{\AA}$ \cite{N2exp}. Fig.~\ref{fig:Ecorr} shows the NNCI results obtained for the three different MO bases and the convergence of $E_\text{corr}$ as a function of the Hilbert space dimensionality. In the plot, we compare our results to the FCI benchmark\cite{ROSSI1999} which was obtained using approximately $10^{10}$ SDets with 34 MOs (black dashed line). This comparison indicates a good performance of the NNCI procedure.
The most remarkable effect, however, is the dramatic improvement we obtain upon increasing the configuration space dimension by including more MOs in the calculation. For 52 MOs (corresponding to a configuration space dimensionality on the order of $10^{12}$), we manage to cross the benchmark line by $\approx 2.6\cdot10^{-3}$ Hartree with less than $8\cdot10^5$ SDets.
This highlights that, while the exponential growth of the configuration space poses technical challenges, the NNCI approach not only remains efficient but actually thrives in higher-dimensional spaces, leveraging the increased number of MOs to achieve superior accuracy with a dramatically reduced number of selected determinants.

Finally, we should also stress that our results indicate that the FCI on 34 MOs from Ref.~\citenum{ROSSI1999} is not converged concerning the number of MOs. Our own NNCI calculations, in turn, still show linear convergence of the energy on the logarithmic $x$-axis. Performing a full FCI calculation with 52 MOs to estimate the convergence limit, however, is beyond reach due to the immense computational demands. It is also worth noting that, while the 34-MO curve in our results should eventually approach the benchmark value reported by Rossi \textit{et al.}\cite{ROSSI1999}, such comparison is not strictly possible because of a different MO basis used.

The convergence for the binding energy and the dissociation curve, however, is different as we will now see. 

\paragraph{N$_2$ Binding energy.} As the next result, we present the N$_2$ binding energy for which we again assume that $E_\text{gs}(a\rightarrow\infty)\approx E_\text{gs}(a=3\,\text{\AA})$ in Eq.~\eqref{eq:Ebind}. 
Fig.~\ref{fig:Ebind_diss}(a) shows the convergence of the NNCI results for three different choices of the size of the underlying MO basis. The values on the $x$-axis correspond to the mean number of Hilbert space dimensions needed to compute $E_\text{gs}(a_\text{min})$ and $E_\text{gs}(a=3\,\text{\AA})$. Due to the clear definition of the spin state of the dissociated molecule, comparison to experimental data at $E_\text{bind}=0.3638\,$ Hartree (see LeRoy \textit{et al.}\citenum{N2exp}) which we show as a black solid line, is straightforward. We also show binding energies of FCI calculations that we estimate from Refs. \citenum{Herzog2023} (18 MOs) and \citenum{Coe_JChemTC_2019} (28 MOs). While we see monotonous convergence toward the experimental value upon increasing the size of the MO basis, the evolution with the NN extension steps shows quite different behavior. For smaller numbers of SDets, the binding energy is overestimated but converges with increased Hilbert space dimensionality to a slightly underestimated value due to different convergence of $E_\text{gs}(a_\text{min})$ and $E_\text{gs}(a=3\,\text{\AA})$. As can be seen from the plot, the converged NNCI results are closer to the experimental results than the values we estimate from the FCI dissociation curves obtained with smaller numbers of MOs. The remarkable fact about this is that the actual number of SDets needed in the NNCI is orders of magnitude smaller than the FCI bases. We can draw the same conclusion as from the correlation energy calculation: The NNCI algorithm benefits greatly from working in very large configuration spaces. 

\begin{table}[t]
\centering
\renewcommand{\arraystretch}{1.2} 
\setlength{\tabcolsep}{12pt}

\begin{tabular}{lc}
\hline
Method & $a_\text{min}$ [Å] \\ \hline
FCI 18 MOs \cite{Herzog2023}   & $1.131$ \\
FCI 28 MOs \cite{Coe_JChemTC_2019}   & $1.117$ \\
NNCI 34 MOs   & $1.107$ \\
NNCI 46 MOs   & $1.103$ \\
NNCI 52 MOs   & $1.104$ \\
Experiment \cite{N2exp}  & 1.098                          \\
\hline
\end{tabular}

\caption{
Comparison of $a_\mathrm{min}$ values from different calculations and experiment. The values were obtained by fitting a modified Morse+Lennard-Jones potential\cite{N2exp}. The fitting error is within $1$ to $4$ m{\AA} (not including errors of the underlying calculation).
}
\label{tab:amin}
\end{table}

\begin{table*}[t]
\centering
\renewcommand{\arraystretch}{1.2} 
\setlength{\tabcolsep}{12pt} 
\begin{tabular*}{\textwidth}{@{\extracolsep{\fill}}lcccc|cc@{}}
\toprule
 & $(N_\mathrm{el},N_\mathrm{orb})$ & $\text{dim}{\cal H}^\text{full}$ & $\text{dim}{\cal H}^\text{NNCI}$ & $E_\mathrm{corr}$ & $E^\text{FCI}_\mathrm{corr}$ & $\text{dim}{\cal H}^\text{FCI}_\text{ref.}$ \\
\midrule
N$_2$   & (10,52)  & $6.75 \cdot 10^{12}$ & $7.16 \cdot 10^{5}$ & -0.349 & -0.346 \cite{ROSSI1999} &  $9.68 \cdot 10^{10}$\\
CO      & (10,46)  & $1.88 \cdot 10^{12}$ & $2.44 \cdot 10^{5}$ & -0.306 & -0.215\textsuperscript{a} & $1.01 \cdot 10^{9}$ \\
NH$_3$  & (8,56)  & $4.49 \cdot 10^{10}$ & $1.93 \cdot 10^{5}$ & -0.221 & -0.208\textsuperscript{a} & $1.41 \cdot 10^{10}$ \\
H$_2$O  & (8,43)  & $1.52 \cdot 10^{10}$ & $2.51 \cdot 10^{5}$ & -0.218 & -0.216\textsuperscript{a} & $1.81 \cdot 10^{9}$ \\
\bottomrule
\end{tabular*}

\caption{Correlation energy $E_\text{corr}$ in Hartree, dimension of full (spin restricted) and selected Hilbert spaces for several molecules with $N_\mathrm{el}$ electrons and $N_\mathrm{orb}$ MOs. Reference values are given for comparison.
}
\label{tab:corr_energy}
\raggedright
\vspace{5pt}
\textsuperscript{a} These calculations\cite{Gao2024} do not use the frozen core approximation (see sec.\ \ref{sec:hf}).
\end{table*}

\paragraph{N$_2$ Dissociation curve -} 
We now turn to the full dissociation curve $E_\text{gs}(a)$ of N$_2$ and compare to an experimental reference curve based on a modified Morse+Lennard-Jones potential fitted to spectroscopic data \cite{N2exp}, which includes the experimentally measured frequencies of the first 20 vibrational states (the highest vibrational state taken into account is roughly centered with respect to the depth of the well).
Fig.~\ref{fig:Ebind_diss}(b) shows data up to $a=3.0$\,\AA\ including results for three different choices for the MO basis as well as the spin-restricted Hartree-Fock curve and two FCI curves which we previously used for the estimate of the binding energy and which are taken from Refs. \citenum{Herzog2023} (based on 18 MOs) and \citenum{Coe_JChemTC_2019} (based on 28 MOs) - please note that the actual data range from the reference publications is plotted as a solid line, while the dashed tail (which is also used for the binding energy estimate) is extrapolated by a fit of the model used for the experimental data.

Overall, the NNCI curves show good agreement with the experimental curve and monotonous convergence toward the experimental data with an increasing number of MOs (as previously seen for $E_\text{bind}$). While $E_\text{bind}$ converges quicker than the full $E_\text{gs}(a)$ curve due to error cancellation, Fig.~\ref{fig:Ebind_diss}(b) demonstrates good agreement of the entire curve from the equilibrium region to the dissociation limit $a\rightarrow \infty$. Comparison to the FCI curves indicates, once more, the benefit of working in high-dimensional spaces in NNCI rather than working with full Hilbert spaces over a smaller number of MOs.

The inset in Fig.~\ref{fig:Ebind_diss}(b) shows a zoom around the minimum of the curve, $a_\text{min}$, which demonstrates that the NNCI calculations improve the HF value considerably toward the experimental value of $a_\text{min}=1.098 \,\text{\AA}$ \cite{N2exp}. A fit of the FCI results for 18 and 28 MOs, and the NNCI results for 34, 46, and 52 MOs to the model used in Ref.~\citenum{N2exp} yields the values for $a_\text{min}$ collected in Tab.~\ref{tab:amin}, highlighting the superior convergence of the equilibrium distance with respect to the number of MOs included in the CI calculation.

\paragraph{H$_2$O, NH$_3$, and CO Correlation energy -}
To demonstrate the transferability of our method further, we have computed the correlation energy for other molecules. 
Tab.~\ref{tab:corr_energy}, shows the correlation energy, $E_\mathrm{corr}$, for H$_2$O, NH$_3$, and CO computed for the same molecular geometry as in Ref.~\citenum{Gao2024} and a comparison with the published results. 
The number of MOs used for the calculations is also shown as well as the number of determinants/dimensionality of the corresponding Hilbert space.
We note that this comparison needs to be taken with some care and is generally not straightforward. Due to the interplay of different error sources, it is challenging to disentangle algorithmic performance from model parameter effects in different ML-supported CI schemes. 
Most notably, the number of MOs included in the calculation has a significant impact on the results. As we have shown in the previous sections, the efficiency of the NNCI-selected bases improves dramatically in higher-dimensional spaces (i.e. larger number of MOs -- see Fig.~\ref{fig:Ecorr}). Unfortunately, FCI data for such large calculations do not exist, to the best of our knowledge, and we can only compare absolute values to smaller calculations.

Nevertheless, the results again clearly highlight the efficiency of our strategy of increasing the number of MOs beyond what is feasible to treat in FCI calculations when selecting SDets with our NNCI method. In this way, we obtain correlation energy values closer to the exact nonrelativistic values than published FCI results with 4-5 orders of magnitude smaller Hilbert space dimensionality.

\section{Summary \& Conclusion}
In summary, we present neural-network-supported selective CI calculations for molecules by integrating Hartree-Fock calculations with the formulation of a cluster Hamiltonian and the application of an NN classifier for selective optimization of the many-body basis. The PAW formalism is used to represent the effect of inner electrons and the smooth pseudo wave functions for the valence electrons are described using a real space grid and/or plane wave basis set, thereby paving the way for applications to larger systems through an embedding approach. This work represents the first application of the SOLAX code \cite{SOLAX} and its combination with a general-purpose electronic structure software package, namely GPAW \cite{GPAW2024}.

For N$_2$, we successfully reproduce and surpass the FCI benchmark correlation energy reported by Rossi \textit{et al.} \cite{ROSSI1999}, achieving this with five orders of magnitude fewer Slater determinants. 
Beyond the correlation energy, we also present the binding energy and the full dissociation curve for N$_2$, finding very good agreement with the experimental potential energy function \cite{N2exp}. 
These results demonstrate the method's ability to capture the critical features of molecular electronic structure with remarkable efficiency.

Evaluation of the (equilibrium) correlation energy for the additional cases NH$_3$, H$_2$O, and CO, shows that the efficient NNCI bases allow us to go beyond the accuracy of FCI calculations based on smaller number of MOs. 
This indicates that NNCI thrives on high dimensional configuration spaces to optimize the many-body ground state without the need to consider the full Hilbert space explicitly.
More subtly, the results also underline the crucial (and well-known) impact of the choice of the underlying single-particle MO basis on the accuracy and efficiency of CI calculations. Therefore, further improvement of the methodology could be obtained by using MO optimization, such as the correlation-optimized virtual orbitals (COVO) \cite{COVO} or, in fact, a procedure that takes feedback from the NNCI calculation to optimize the underlying MO basis with the goal to expand the ground state on the smallest possible number of determinants.

Future applications of the NNCI scheme include geometry optimization of atomic structures and calculations of the energy and intensity of optical transitions, where recently developed state-specific mean-field methods \cite{Levi2020} can provide a basis of orbitals optimized for excited electronic states.

While the results for our and related strategies are promising, NN-supported approaches to electronic structure calculations are still relatively new. Future benchmarks against well-established non-ML methods, such as FCIQMC \cite{FCIQMC}, are therefore necessary. Such comparisons will help clarify questions regarding the general convergence behavior of these methods and their potential for further optimization. 


\section{Appendix}
\subsection{Calculation of single- and two-particle integrals in the projector augmented wave approach}

Within the PAW formalism, the so-called ``all-electron'' wave functions, which contain cusps at the positions of the nuclei, are written as
\begin{equation}
 \psi_{i,\sigma}(\mathbf{r}) = \hat{\mathcal{T}}\tilde{\psi}_{i,\sigma}(\mathbf{r})
\end{equation}
where $\tilde{\psi}_{i,\sigma}$ are ``pseudo-electron'' wave functions, which are smooth everywhere. $\hat{\mathcal{T}}$ is a linear transformation operator, which corrects for the smooth description of the electronic wave functions near the positions of the nuclei 
\begin{equation}
 \hat{\mathcal{T}} = 1 + \sum_{a\alpha}(\ket{\varphi^a_\alpha} - \ket{\tilde{\varphi}^a_\alpha})\bra{\tilde{p}^a_\alpha}
\end{equation}
Here, $\varphi^a_\alpha$ and $\tilde{\varphi}^a_\alpha$ are partial waves describing the all-electron and pseudo-electron wave functions in an atomic region of radius $r^a_c$ around each nucleus $a$.
The all-electron and pseudo-electron partial waves are required to be identical beyond the radius $r^a_c$, i.e. $\varphi^a_\alpha(\mathbf{r})=\tilde{\varphi}^a_\alpha(\mathbf{r})$ for $\mid\mathbf{r} - \mathbf{R}^a\mid > r^a_c$.
$\tilde{p}^a_\alpha$ are smooth projection functions, which satisfy 
\begin{align}
 \sum_\alpha \ket{\tilde{\varphi}^a_\alpha} \bra{\tilde{p}^a_\alpha} =&\, 1 \ \ \ \ \text{for} \mid\mathbf{r} - \mathbf{R}^a\mid \leq r^a_c \\
 \braket{\tilde{p}^a_\alpha| \tilde{\phi}^a_\beta} =&\, \delta_{\alpha\beta} \ \ \ \ \text{for} \mid\mathbf{r} - \mathbf{R}^a\mid \leq r^a_c
\end{align}
such that
\begin{align}
 \ket{\tilde{\psi}_{i,\sigma}} =& \sum_\alpha \ket{\tilde{\varphi}^a_\alpha} P^a_{\alpha i,\sigma} \ \ \ \ \text{for} \mid\mathbf{r} - \mathbf{R}^a\mid \leq r^a_c \\
 \ket{\psi_{i,\sigma}} =& \sum_\alpha\ket{\varphi^a_\alpha} P^a_{\alpha i,\sigma} \ \ \ \ \text{for} \mid\mathbf{r} - \mathbf{R}^a\mid \leq r^a_c
\end{align}
where $P^a_{\alpha i,\sigma} = \braket{\tilde{p}^a_\alpha| \tilde{\psi}_{i,\sigma}}$, i.e. the all- and pseudo-electron wave functions can be expanded into partial waves with the same linear expansion coefficients. It is convenient to define an atomic density matrix, which for a given state is
\begin{equation}
 D^a_{\alpha\beta i,\sigma} = P^{a*}_{\alpha i,\sigma}P^a_{\beta i,\sigma}
\end{equation}
or similarly for any pair of states
\begin{equation}
 D^a_{\alpha\beta ij,\sigma} = P^{a*}_{\alpha i,\sigma}P^a_{\beta j,\sigma}
\end{equation}
For clarity, the spin index is suppressed in the following sections.


\subsubsection{Single particle integrals and PAW}
\label{app:HF_single}

The single-particle integrals in the PAW formalism are
\begin{align}
 t_{ij} &= \int \text{d}\mathbf{r}\, \psi^*_i(\mathbf{r}) \left(-\frac{1}{2}\nabla^2 + V^\text{eff}(\mathbf{r}) \right) \psi_j(\mathbf{r}) \nonumber \\
 &= -\int \mathrm{d}\mathbf{r}\,
 \tilde{\psi}_i(\mathbf{r})\frac{\nabla^2}{2}\tilde{\psi}_j(\mathbf{r})
 \nonumber \\
 &+ \int \mathrm{d}\mathbf{r} \, \tilde{n}_{ij}(\mathbf{r})\left(\tilde{V}_\mathrm{Coul}(\mathbf{r}) + \tilde{V}_\mathrm{XX}(\mathbf{r}) \right) \nonumber \\
 &+ \left(\frac{\partial\Delta E^a}{\partial D^a_{\gamma\delta ij}}
  + \int \mathrm{d}\mathbf{r}\, \tilde{V}_\mathrm{Coul}(\mathbf{r})\frac{\partial \tilde{\rho}(\mathbf{r})}{\partial D^a_\mathrm{\gamma\delta}}\right)D^a_{\alpha\beta ij}\,,
\end{align}
where $\tilde{V}_\mathrm{Coul}$, $\tilde{V}_\mathrm{XX}$ are the Coulomb and exchange potentials formed by the pseudo electron-wave functions, respectively, $D^a_\mathrm{\gamma\delta} = \sum_i D^a_{\alpha\beta i,\sigma} = \sum_i P^{a*}_{\alpha i,\sigma}P^a_{\beta i,\sigma}$, $\tilde{\rho}(\mathbf{r})$ is a pseudo total charge density obtained by adding compensation charges for the nuclei to the pseudo-electron density, and $\Delta E^a$ are atomic corrections to the kinetic, Coulomb and exchange energy, as defined in Refs. \citenum{GPAW2024}, ~\citenum{Enkovaara2010}, and ~\citenum{PhysRevB.71.035109}. The indexation $ij$ is over the spatial electronic wave functions.


\subsubsection{Two particle integrals and PAW}
\label{app:HF_pair}

The elements of the interaction tensor are

\begin{align}\label{eq:exchange-pair}
U_{ijkl} = & \int \mathrm{d}\mathbf{r} \mathrm{d}\mathbf{r}'\,
           \frac{\psi_{i}^{*}(\mathbf{r})\psi_{j}(\mathbf{r})
           \psi^*_{k}(\mathbf{r}')\psi_{l}(\mathbf{r}')}{\mid \mathbf{r} - \mathbf{r}' \mid} \\
        = & \int \mathrm{d}\mathbf{r} \mathrm{d}\mathbf{r}'\,
           \frac{n_{ij}(\mathbf{r})n_{kl}(\mathbf{r}')}{\mid \mathbf{r} - \mathbf{r}' \mid}\,,\nonumber
\end{align}
where, for instance, $U_{iijj}$ and $U_{ijij}$ are Coulomb and exchange integrals, respectively. Pair valence orbital densities $n_{ij}(\mathbf{r})$ are given by
\begin{align}
 n_{ij}(\mathbf{r}) & =  \braket{\psi_{i}|\mathbf{r}}\braket{\mathbf{r}|\psi_{j}} \nonumber \\
 & = \tilde{n}_{ij}(\mathbf{r}) + \sum_a(n^a_{ij}(\mathbf{r}^l) - \tilde{n}^a_{ij}(\mathbf{r}^l))
\end{align}
where $\tilde{n}_{ij}(\mathbf{r}) = \braket{\tilde{\psi}_{i}|\mathbf{r}}\braket{\mathbf{r}|\tilde{\psi}_{j}}$ are pseudo pair orbital densities, and $n^a_{ij}$ and $\tilde{n}^a_{ij}$ are atom-centered pair densities:
\begin{align}
n^a_{ij}(\mathbf{r}^l)&= \sum_{\alpha\beta}D^a_{\alpha\beta ij,\sigma}\braket{\varphi_{\alpha}^{a}|\mathbf{r}^l}\braket{\mathbf{r}^l|\varphi_{\beta}^{a}} \\
\tilde{n}^a_{ij}(\mathbf{r}^l)&=\sum_{\alpha\beta}D^a_{\alpha\beta ij,\sigma}\braket{\tilde{\varphi}_{\alpha}^{a}|\mathbf{r}^l}\braket{\mathbf{r}^l|\tilde{\varphi}_{\beta}^{a}}
\end{align}
All atom-centered partial waves and densities are represented on a radial grid, as indicated by $\mathbf{r}^l$. 
The pseudo pair densities and atom-centered pseudo pair densities are modified by adding and subtracting atom-centered compensation charges $\tilde{\mathcal{Z}}^a_{ij}$:
\begin{align}
 n_{ij}(\mathbf{r}) & = \tilde{n}_{ij}(\mathbf{r}) + \sum_a\tilde{\mathcal{Z}}^a_{ij}(\mathbf{r}) \nonumber \\ 
 & + \sum_a(n^a_{ij}(\mathbf{r}^l) - \tilde{n}^a_{ij}(\mathbf{r}^l) - \tilde{\mathcal{Z}}^a_{ij}(\mathbf{r}^l)) \label{eq:pair-compensated}
\end{align}
The $\tilde{\mathcal{Z}}^a_{ij}$ are chosen such that $n^a_{ij}(\mathbf{r}^l) - \tilde{n}^a_{ij}(\mathbf{r}^l) - \tilde{\mathcal{Z}}^a_{ij}(\mathbf{r}^l)$ have vanishing multipole moments. This is achieved by expanding the compensation charges in terms of real-space solid harmonics, ensuring that the atomic regions are electrostatically decoupled up to and including the quadrupole moment~\cite{Enkovaara2010, PhysRevB.71.035109}.
 
With the electrostatic decoupling of atomic regions, expanding eq.~\ref{eq:exchange-pair}
in terms of eq.~\ref{eq:pair-compensated} gives
\begin{align}
U_{ijkl} = & \int \mathrm{d}\mathbf{r} \mathrm{d}\mathbf{r}'\,
\frac{\tilde{\rho}_{ij}(\mathbf{r}) \tilde{\rho}_{kl}(\mathbf{r}')
      }
      {\left|\mathbf{r} - \mathbf{r}' \right|}  \nonumber \\
+& \sum_a\int \mathrm{d}\mathbf{r}^l \mathrm{d}\mathbf{r}^{l'}\,
\frac{\Delta \rho^a_{ij}(\mathbf{r}^l)
\Delta \rho^a_{kl}(\mathbf{r}^{l'})}
{\left| \mathbf{r}^l - \mathbf{r}^{l'} \right|}
\end{align}
where $\tilde{\rho}_{ij}(\mathbf{r})=\tilde{n}_{ij}(\mathbf{r}) + \sum_a\tilde{\mathcal{Z}}^a_{ij}(\mathbf{r})$, and $\Delta \rho^a_{ij}(\mathbf{r}^l)=n^a_{ij}(\mathbf{r}^l) - \tilde{n}^a_{ij}(\mathbf{r}^l) - \tilde{\mathcal{Z}}^a_{ij}(\mathbf{r}^l)$. Each element of the interaction tensor conveniently separates into a pseudo and an atomic correction part
\begin{equation}
U_{ijkl} = \tilde{U}_{ijkl} + \sum_a\Delta U^a_{ijkl}
\end{equation}
It can be shown \cite{PhysRevB.71.035109} that
\begin{equation}
 \Delta U^a_{ijkl} = 
 2\sum_{\alpha\beta\gamma\delta}D^a_{\alpha\beta ij}
 C^a_{\alpha\beta\gamma\delta}D^a_{\gamma\delta kl}
\end{equation}
where the atomic Coulomb tensor $C^a_{\alpha\beta\gamma\delta}$ is precalculated and stored.

For the pseudo part, we define the pseudo pair density potential
\begin{equation}
 \tilde{\Phi}_{ij}(\mathbf{r}) = \int \mathrm{d}\mathbf{r}'\,\frac{\tilde{\rho}_{ij}(\mathbf{r}')}{\left| \mathbf{r} - \mathbf{r}' \right|}\,,
\end{equation}
which is solved by using standard Poisson solvers. 
Thus, the pseudo term is
\begin{equation}
 \tilde{U}_{ijkl} = \int \mathrm{d}\mathbf{r}\;\tilde{\Phi}_{ij}(\mathbf{r})\tilde{\rho}_{kl}(\mathbf{r})
\end{equation}
For real-valued functions, the indices $ijkl$ are invariant according to the following symmetry operations $i\leftrightarrow j, \ \ k \leftrightarrow l, \ \ ij\leftrightarrow kl$, and any combination thereof, leading to an eightfold reduction in the number of terms that need to be evaluated.
Note that $\tilde{n}_{ii}$ contain a net monopole ($\int \mathrm{d}V \tilde{n}_{ii} \neq 0$). In the case of a plane wave basis set, a charge-neutralizing background is added to the simulation cell (with constant value $\frac{1}{V}\int \mathrm{d}V \tilde{n}_{ii}$), to obtain a charge-neutral electrostatic potential. For all terms involving the symmetry $i=j \wedge k=l$, the shift in energy due to the charge neutralization is corrected afterward.


\subsubsection{Double counting correction}
\label{app:DC}
To avoid double counting the mean-field potential of the Hartree-Fock solution in the full CI Hamiltonian, we subtract a correction term. For a Hartree-Fock (in contrast to a density functional theory) starting point, this term is well-defined and corresponds to the mean-field decoupled interaction operator

\begin{align}
\mathrm{MF}\left[H^\mathrm{int}\right] 
&= \sum_{\substack{i,j,k,l \\ \sigma,\sigma'}} U_{ijkl} \Big( \braket{ c^\dagger_{i,\sigma} c^{\phantom{\dagger}}_{j,\sigma} } \, c^\dagger_{k,\sigma'} c^{\phantom{\dagger}}_{l,\sigma'} \nonumber \\ 
&\hspace{2cm} + c^\dagger_{i,\sigma} c^{\phantom{\dagger}}_{j,\sigma} \, \braket{ c^\dagger_{k,\sigma'} c^{\phantom{\dagger}}_{l,\sigma'} } \nonumber \\ 
&\hspace{2cm} - \braket{ c^\dagger_{i,\sigma} c^{\phantom{\dagger}}_{j,\sigma} } \, \braket{ c^\dagger_{k,\sigma'} c^{\phantom{\dagger}}_{l,\sigma'} } \Big)
\end{align}
where the expectation values $\braket{ c^\dagger_{i,\sigma} c^{\phantom{\dagger}}_{j,\sigma} }$ are matrix elements of the Hartree-Fock density matrix.   

\subsection{Neural network architecture and training\label{app:NN}}

For the basis state selection procedure, an NN classifier of the convolutional type \cite{Goodfellow2016} is used, see Fig.~\ref{fig:nn_architecture}. The NN receives as input (A) a candidate SDet 
$\ket{\phi}$ in occupation number representation, $i.e.$ a string consisting of 0 and 1.
The input is split into two spin channels and then passes through a filter kernel of size 2 (B), generating 64 feature maps (C). These are subsequently processed by a kernel of size 1, resulting in 4 output channels (D). The latter is flattened and forwarded to a dense block (E) which ends with an output layer consisting of two neurons (F). These neurons classify the input SDet 
as ``important" or ``unimportant" by applying a softmax activation function. The latter ensures that the outputs lie between 0 and 1 and sum up to 1, and are therefore interpretable as the corresponding probabilities.

In all hidden NN layers, the rectified linear unit (ReLU) is employed as the activation function, and the network’s performance is evaluated using categorical cross-entropy. The Adam algorithm \cite{kingma2017adam} is used for training, which terminates after no improvement is observed over three consecutive epochs, a method known as ``early stopping with patience". For further comprehensive details of the iterative algorithm and computational protocol, we refer to Ref.~\citenum{Bilous2024}.


\begin{acknowledgement}
The authors thank Hong Gao for providing the correlation energy for their FCI calculations\cite{Gao2024} on the N$_2$, CO, H$_2$O, and NH$_3$ molecules. This work was supported by the Icelandic Research Fund (grant agreements nos.\ 217751, 217734) and the University of Iceland Research Fund. Computer resources, data storage, and user support were provided by the 
Icelandic Research e-Infrastructure (IREI), funded by the Icelandic Infrastructure Fund. The authors further gratefully acknowledge the scientific support and HPC resources provided by the Erlangen National High Performance Computing Center (NHR@FAU) of the Friedrich-Alexander-Universität Erlangen-Nürnberg (FAU). PB gratefully acknowledges the ARTEMIS funding via the QuantERA program of the European Union provided by German Federal Ministry of Education and Research under the grant 13N16360 within the program ``From basic research to market''.
\end{acknowledgement}


\vskip 0.6 true cm

{\bf  Data and Software Availability}
Data related to the results presented in this article and instructions on the generation thereof are available at Zenodo.\cite{ZenodoData} 

\bibliography{main_references}

\end{document}